\documentclass[fleqn,10pt,twocolumn]{AROB-ISBC-SWARM23}

\usepackage{algorithm}
\usepackage[noend]{algpseudocode}

\usepackage{here}
\newcommand{\figurename}{Fig.}
\newcommand{\tablename}{Table}
\def\fgref#1{\figurename\ref{#1}}
\def\tbref#1{\tablename\ref{#1}}

\title{Generating corneal panoramic images from
contact specular microscope images }

\author{Yusuke Nagira${}^{1\dagger}$, Yuzuha Hara${}^{2}$, Satoru Hiwa${}^{2}$\\
Naoki Okumura${}^{3}$, Noriko Koizumi${}^{3}$ and Tomoyuki Hiroyasu${}^{2}$}
\speaker{Yusuke Nagira}

\affils{${}^{1}$Graduate School of Life and Medical Sciences, Doshisha University, Japan\\
${}^{2}$Department of Biomedical Sciences and Informatics, Doshisha University, Japan\\
${}^{3}$Department of Biomedical Engineering, Faculty of Life and Medical Sciences, Doshisha University, Japan\\
(Tel: +81-774-65-6020, E-mail: tomo@is.doshisha.ac.jp)\\
}
\abstract{%
 The contact specular microscope has a wider angle of view than that of the non-contact specular microscope but still cannot capture an image of the entire cornea. To obtain such an image, it is necessary to prepare film on the parts of the image captured sequentially and combine them to create a complete image. This study proposes a framework to automatically generate an entire corneal image from videos captured using a contact specular microscope. Relatively focused images were extracted from the videos and panoramic compositing was performed. If an entire image can be generated, it is possible to detect guttae from the image and examine the extent of their presence. The system was implemented and the effectiveness of the proposed framework was examined. The system was implemented using custom-made composite software, Image Composite Software (ICS, K.I. Technology Co., Ltd., Japan, internal algorithms not disclosed), and a supervised learning model using U-Net was used for guttae detection. Several images were correctly synthesized when the constructed system was applied to 94 different corneal videos obtained from Fuchs endothelial corneal dystrophy (FECD) mouse model. The implementation and application of the method to the data in this study confirmed its effectiveness. Owing to the minimal quantitative evaluation performed, such as accuracy with implementation, it may pose some limitations for future investigations.
}

\keywords{%
U-Net, Semantic Segmentation, Fuchs Endothelial Corneal Dystrophy, Corneal Endothelial Cell
}

\begin{document}

\maketitle

%-----------------------------------------------------------------------

\section{Introduction}
Fuchs endothelial corneal dystrophy (FECD) is a bilateral disease, wherein corneal endothelial cells are unable to maintain their hexagonal shape. It is characterized by the accelerated loss of corneal endothelial cells with changes in Descemet’s membrane, resulting in the formation of an extracellular matrix called guttae \cite{eghrari2015fuchs}.
In the United States, it is estimated that $4\%$ of people over 40 years of age are affected by the disease, occurring more commonly in women and more frequently in people in their 40s and 50s. Corneal endothelial pump function is lost as the disease progresses, causing corneal edema \cite{nanda2019current}.
Presently, corneal transplantation is the only reliable treatment, and FECD accounts for $39\%$ of all corneal transplants performed, making it the most common cause of corneal transplantation worldwide \cite{gain2016global}.

Rho kinase inhibitors have been reported to promote cell proliferation and adhesion to substrates, inhibit corneal endothelial cell apoptosis, and promote wound healing. Therefore, using Rho kinase inhibitor eye drops is a potential novel treatment approach alternative to corneal transplantation \cite{okumura2017application}\cite{okumura2017role}.

In drug discovery research for FECD, the state of the corneal endothelium, such as the guttae, is observed before and after the drug use and evaluated based on the increase or decrease in the number of cells. Doctors and researchers widely use specular microscopes to observe the state of the corneal endothelium. However, the range of the microscope’s imaging capability is limited and the current practice estimates the state of the entire cornea from its center. The endothelial cell density (ECD) is essential for understanding the pathogenesis of FECD. However, ECD cannot be measured accurately owing to the presence of guttae; hence, the number of cells is measured manually \cite{mclaren2014objective}.

Using a mouse model to demonstrate the pathogenesis of FECD, studies have been conducted on the segmentation of guttae using U-Net and the calculation of cell density in areas excluding guttae \cite{okumura2021u}\cite{nishikawa2021deep}.
In previous studies on the panoramic synthesis of the corneal endothelium, focused images were extracted manually and stitched together. Panoramic compositing of the entire cornea is yet to be performed because the images were localized panoramic images and did not represent the entire cornea \cite{tanaka2017panoramic}.
Conventional panorama synthesis software such as AutoStitch \cite{brown2007automatic} does not consider the order of the input images used for synthesis. Instead, it extracts the image features using Scale-Invariant Feature Transform (SIFT) \cite{lowe2004distinctive} and stitches the matching features together. When pasting an image, it is deformed and scaled. For images with similar characteristics, such as corneal endothelial cells, the position of the image to be pasted may be incorrect or the shape of the cells may be deformed owing to the deformation of the image. In this case, accurate values cannot be obtained when calculating the area of the cells or the percentage of guttae. Therefore, the original image needed to have as minimal deformations as possible in the combined image. Alternatively, it is necessary to provide a mechanism to access the original version of the image of interest.

This study proposes a framework for a system that generates images of the entire corneal endothelium from videos obtained using a contact specular microscope. In the proposed framework, the focused images are extracted from the video images, feature extraction is performed, and the images are synthesized. During synthesis, the system reduces or deforms the original image to the minimum and adds a feature that allows access to the original image of the area of interest. This study examined the proposed framework by building a system using two implementation methods. Further, we added a function for automatically detecting guttae using U-Net, a type of Deep Learning. This study presents a proposed framework and an example of its implementation. Quantitative evaluation of whole corneal endothelial images and gut detection is insufficient, which poses a limitation that must be addressed in future studies.

\begin{figure*}[t]
\begin{center}
\includegraphics[width=2\columnwidth]{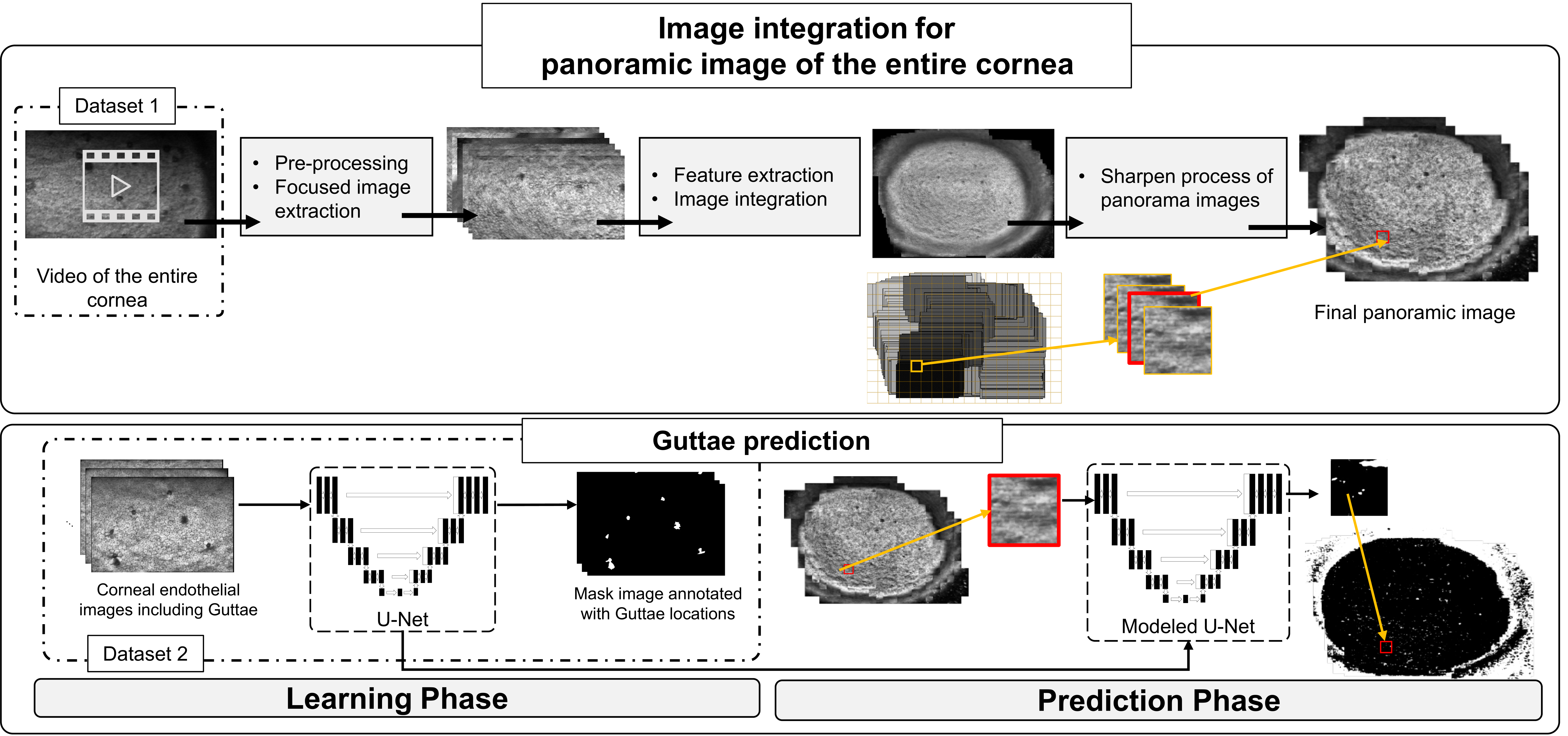}
\caption{\label{framework_overview} Overview of the proposed framework}
\end{center}
\end{figure*}

\section{Framework of corneal panoramic image generation from contact specular microscope images}
\subsection{Overview of the proposed framework}
\fgref{framework_overview} presents an overview of the proposed framework. An image of the entire cornea was generated from a video frame of the cornea captured using a contact-type specular microscope. First, a focused still image was extracted from the target video image (dataset 1). Second, the features of still images were extracted. Third, the matching feature points were combined to create a panoramic image. Next, the panoramic image was divided into grid regions and the most focused image was selected from each region. Guttae were detected in the combined panoramic images using deep learning. Additionally, still images of the corneal endothelium containing the guttae and mask images exhibiting the location of the guttae were used for the model generation (dataset 2).

\subsection{Extraction of in-focus images}
A group of still images was extracted from the captured videos. The entire corneal endothelium was captured and converted to a single frame-by-frame image. If there were $N$ frames in the video, $N$ images were the output in total that were then divided into five groups in chronological order. The image with the highest in-focus evaluation index was selected from each group.

\subsection{Creating panoramic images}
The algorithm for generating a single panoramic image is accomplished through the following steps. First, characteristic points in the images were extracted. Subsequently, a curve connecting the characteristic points was obtained. Here, the optimal curve connecting the extracted characteristic points was obtained. This curve was then used to enlarge or interpolate the image. A panoramic image was generated. In the following experiments, two algorithms were prepared and the results were compared.

\subsection{Image sharpening process}
The synthesized panoramic images were mostly overlapped images. Additionally, the synthesized image is often blurred because of the transparency and brightness of the image change. Therefore, a method was developed to obtain clearer images. The implementation of the sharpening process is described in the next section.

\subsection{Creating the Guttae Classifier by U-Net}
The U-Net is a neural network commonly used for image segmentation. U-Net uses a convolutional neural network to encode an input image as a feature map, subsequently decoding that feature map to separate specific objects in the input image. It is possible to prepare a dataset of corneal images and train U-Net using this dataset to generate a model for extracting the guttae.

%%%%%%%%%%%%%%%%%%%%%%%%%%%%%%%%%%%%%%%%%%%%%%%%%%%%%%%%%%%
\section{System implementation and data application}
\subsection{Outline}
This study implemented a system to confirm the effectiveness of the proposed framework. Corneal videos obtained from the FECD mouse model were processed to obtain panoramic images of the cornea. Tenengrad was used to obtain the in-focus images. Finally, two different applications were used to generate panoramic images.

\subsection{Mouse Model of FECD}
This study used images of whole corneal endothelial cells from the FECD pathology mouse model. A single nucleotide mutation in COL8A2 generated these genes, and it has been reported that guttae increase over time. The Tissue Engineering Laboratory, Graduate School of Biomedical Sciences, Doshisha University, provided the images. The images were taken using a prototype KSSP slit-scanning wide-field contact specular microscope (Konan Medical, Inc., Nishinomiya, Japan), had a resolution of 1620 × 1080 [pixels] at a frame rate of 29.9 [fps] in MOV file format. The images were taken in the following order: 1) starting from the center of the cornea; 2) moving to the top of the cornea; 3) moving to the left; 4) filming from the top to the bottom of the cornea; 5) moving around the right side; 6) filming from the bottom to the center of the cornea. In this study, 94 videos were prepared and used.

\subsection{Extraction of in-focus images by Tenengrad}
The focus evaluation index was calculated as follows: Tenengrad \cite{krotkov1988focusing}\cite{pertuz2013analysis} value, which is the gradient of the image based on the pixel value, is calculated for each region, where $G_x$ and $G_y$ are the convolved values of the Sobel operator of the pixel values in the x-direction and y-direction, respectively.

\begin{equation}
\Phi_{x,y} = \sum_{(i,j)\in \Omega(x, y)}(G_x(i, j)^2+G_y(i, j)^2)
\end{equation}
The Sobel operator is expressed as
\[
  K_x = \left[
    \begin{array}{rrr}
      -1 & 0 & 1 \\
      -2 & 0 & 2 \\
      -1 & 0 & 1
    \end{array}
  \right],
  K_y = \left[
    \begin{array}{rrr}
      -1 & -2 & -1 \\
      0 & 0 & 0 \\
      1 & 2 & 1
    \end{array}
  \right]
\]

The highest value in the quadratic area of a single image was considered the focal value for that image. When this value was calculated for the entire image, the gradient was smaller in the area containing the edge of the corneal endothelium. In contrast, the gradient increased in the area containing corneal endothelial cells. Thus, as the area of the rim increases, the Tenengrad value for the entire image becomes smaller. This procedure prevents the corneal endothelium from being excluded from the image even if it is appropriately captured.

\subsection{Creating the idealized panoramic artificial CECs image data}
To confirm the effectiveness of the panorama synthesis software, a set of idealized panoramic artificial corneal endothelial cell images were created with no blurring or focus mismatch on the extracted images. These images were obtained using the GNU Image Manipulation Program (GIMP). These artificial images were created based on the synthesis results obtained using the panorama synthesis software described below. A layer was added to the composite image, the cell membrane of the corneal endothelium was traced, and the areas considered to be guttae were painted black. The color of the surrounding endothelial cells was extracted from the layer depicting the cell membrane and guttae using the color picker function. The layers are filled with the same color. This process was applied to the entire cornea to create artificial images. For areas where the cell membrane was not visible owing to issues such as focus mismatch or blurring, the cell membrane was depicted by referring to another image in which the cell membrane could be observed appropriately. Idealized panoramic artificial corneal endothelial cell images were created that mimicked the distribution and size of the guttae, as well as the size and shape of the corneal endothelial cells. The created image was 1870 × 1080 [pixel] in size and was cropped and stored by moving approximately 10 [pixels] from the center to the top, counterclockwise from the top, and counterclockwise from top to bottom, mimicking the movement of a motion camera.

\subsection{Creating panoramic images}
In the composite process, two types of applications were used; the Image Composite Software (ICS) and the panorama synthesis algorithm implemented in OpenCV. The algorithms are explained as follows.

\subsubsection{Image Composite Software (ICS)}
Image Composite Software (K.I. Technology CO., LTD., Yokohama, Kanagawa, Japan) is a panorama compositive software created with specifications suitable for image compositing corneal endothelial images. This application was used for the composite process and was custom-made. Since this is a commercial application, we cannot explain the details of its contents due to copyright. The original image is not reduced or enlarged when the images are superimposed. Additionally, an API to access the original image is provided, allowing quick access to the original image of the area of interest. A flowchart of the panorama compositing process is shown in Algorithm \ref{alg1}, where the first image is used as the reference image, and the regions that match the first image are searched in order of image number. If no match is observed, the image merged with the reference image is used as the reference for the subsequent image, and the process is repeated. The image merging is terminated when ten consecutive images are not observed to match the reference image.

\begin{algorithm}[t]
\begin{algorithmic}[1]
\caption{Image Composite Software processing details}
\label{alg1}
    \State $i$ = 0, $j$ = 0
    \While{$i$ + $j$ + 1 $\leq N$}
    	\State A = Images[$i$] 
        \State $j$ = 0
    	\While{$j \leq 10$}
            \State B = Images[$i$+$j$+1]
    		\If{Find matching area with A}
                \State A = Stitch B onto A
    			\State $i$ = $i$ + 1, $j$ = 0
    		\Else
    			\State $j$ = $j$ + 1
            \EndIf
        \EndWhile
    \EndWhile
\end{algorithmic}
\end{algorithm}

\subsubsection{Panorama synthesis algorithm implemented in OpenCV}
Because the details of the process in ICS are not publicly available, we implemented an algorithm similar to Algorithm \ref{alg2} that mimics the ICS process, using OpenCV, an open-source computer vision library. The $K$th image was the closest to the end of the shooting. The $j$th image and the $i + 1$ image are matched for SIFT features, and if the two images have many similar features, the degree of change between the images is calculated and added to the list. The $i$th image is matched to the $i + 1$ image. If the two images have few feature points and cannot be matched, add 1 to the values of $i$ and f and perform the SIFT feature extraction again. If this process fails ten times, add 1 to the value of $j$ and perform the process again taking $j$ equivalent to $i$, that is, $j = i$. Using the above algorithm, the degree of change in coordinates between the images used for composition and the images can be calculated. The global coordinates of the entire composite coordinates can be obtained by setting the smallest values of the $x$ and $y$ coordinates to zero and calculating the cumulative sum till that point.

\begin{algorithm}[t]
\begin{algorithmic}[1]
\caption{Calculate the difference in coordinates between images}
\label{alg2}
    \State $i$ = 0
    \State coord = [], usedImages = []
    \While{$j \leq N$}
    	\State $j$ = $i$
        \State error = 0
        \State flag = \texttt{True}
    	\While{flag}
            \State Extract SIFT features of Image[$j$] and Image[$i$+1]
    		\If{Two images could be feature matched}
    			\State C = cal\_coord\_diff(Image[$j$], Image[$i$+1])
    			\State coord.append(C)
    			\State usedImages.append(Image[$j$], Image[$i$+1])
    			\State $i$ += 1
                \State flag = \texttt{False}
    		\Else
    			\State error += 1
   			\If{error $\geq$ 10}
    				\State $j$ += 1
    			\Else
    				\State $i$ += 1
                \EndIf
            \EndIf
        \EndWhile
    \EndWhile
    \State usedImages = list(set(usedImages)
    \State \textbf{return} coord, usedImages
    \end{algorithmic}
\end{algorithm}

\subsection{Image sharpening process}
We divided the combined panoramic image into 64 × 64 [pixel] grid regions. The image number of each composite image and the coordinates of the constituent images were obtained from the coordinates of the panoramic image. The coordinates of the constituent images in the upper-left corner of each grid region were obtained. Tenengrad values were calculated for the cropped images. The image with the highest Tenengrad value among the cropped images was pasted onto a newly created blank image of the same size as the panoramic image with the exact extracted coordinates. These processes were performed in all regions. The image with the highest Tenengrad value among the multiple overlapping images was pasted to obtain a clear image.

\subsection{Creating the Guttae Classifier}
\subsubsection{Creating Dataset}
A large amount of data is required for training using networks. However, since the amount of data provided in this study was small, it was necessary to augment the data. Additionally, due to the large size of the image data, it was necessary to reduce the size of the images to use them in the dataset. Because image resizing results in a loss of information, we developed a new data augmentation method for small and large image data. The corneal endothelial cell image and mask image showing the location of the guttae were divided into a grid of 64 × 64 [pixels]. The image was clipped three times by 16 [pixels] to the right and three times by 16 [pixels] to the bottom, thereby shifting the image area. Thus, the images were expanded 16 times for a single-grid region. Images that did not contain guttae were excluded from the dataset.

\subsubsection{Training and Prediction in CNN}
Twenty-one corneal endothelial cell images with a mask image showing the location of the guttae were prepared (\fgref{framework_overview} Dataset 2).
 The Segmentation model PyTorch, which is a Python library for implementing segmentation-specific CNNs, was used in the experiments. The best encoder backbone is determined using ResNet18, ResNet34, ResNet50, VGG11, and VGG16. Twenty-one images were divided into two sets of eighteen and three images; eighteen were used to develop this model and three were used to determine the backbone. The node weights of ImageNet were transferred to this model, and fine-tuning was performed. The network of U-Net \cite{ronneberger2015u} encoders modified to ResNet50 \cite{he2016deep} is the best backbone. Adam \cite{kingma2014adam} was used as the optimization function with an initial learning rate of 1e-4, wherein the loss function is the least squares error. Predictions were made on a 64 × 64 [pixels] image extracted during the sharpening process, pasted to the original position, and the location of the guttae was predicted for the panoramic image.

\section{Results and Discussion}
\begin{figure*}[t]
\begin{center}
\includegraphics[width=2\columnwidth]{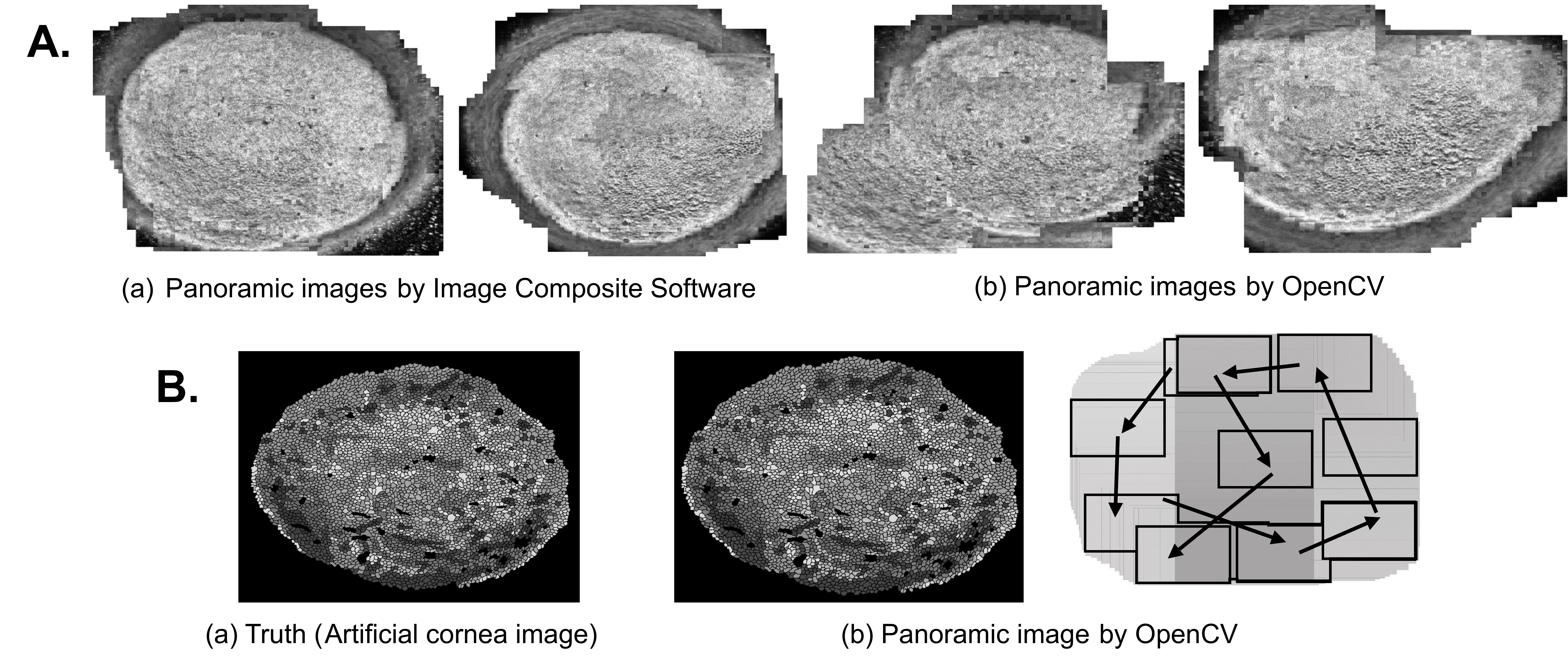}
\caption{\label{result_OpenCV_ICS} A: Comparison of results with Image Composite Software and software implemented in OpenCV. B: Results with OpenCV implemented software on artificial cornea data.(a) shows the composite result and (b) shows the overlap of the component images.}
\end{center}
\end{figure*}
\subsection{Comparison of algorithms implemented in ICS and OpenCV}
An example of a mouse corneal endothelial cell is shown in \fgref{result_OpenCV_ICS}A. The images synthesized using ICS were more circular than those synthesized using OpenCV. This result indicates that the synthesis was performed correctly. On the other hand, images synthesized using the OpenCV algorithm were not circular and were often pasted in incorrect locations.

An example of the synthesis result of the algorithm implemented in OpenCV for artificial cornea data is shown in \fgref{result_OpenCV_ICS}B. This is the synthesis result when the focus is perfectly aligned and there are no blurred images. The result is almost the same as the correct data, where there is no unnatural overlap between the composite images, indicating that compositing was performed correctly. If the in-focus images are well extracted, they can be integrated well using OpenCV. Overall, ICS is a more robust method.

\subsection{Synthesis results with ICS}
\begin{figure*}[t]
\begin{center}
\includegraphics[width=2\columnwidth]{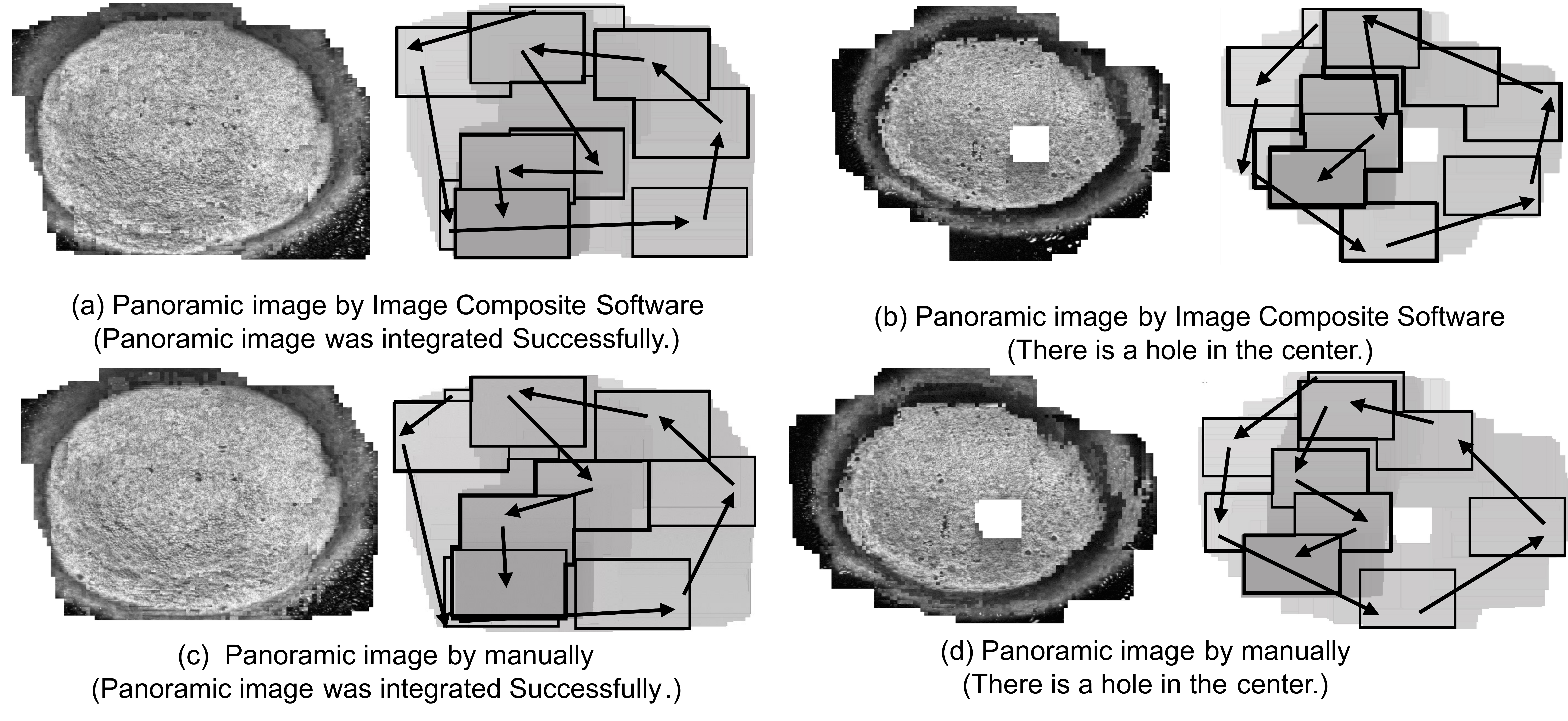}
\caption{\label{result_ics} Integrated image results by Image Composite Software and manual composite.The left side of each represents the composite result, and the right side represents the overlap of the component images.}
\end{center}
\end{figure*}
As previously mentioned, when taking moving images of the mouse cornea with the contact specular microscope, the images are taken in an upward direction from the center of the cornea, then leftward along the edge of the cornea, and downward once around the edge. Next, it was photographed clockwise and then down to the center. For the 94 videos, the left- and right-rotated portions were split, and for each case, an integrated image was created using the ICS. The results are presented in \tbref{tb:result_ICS}.
An image was classified as "image dropout" if the center of the image was missing, "distorted shape" if the shape was distorted instead of being circular, and "unnatural paste" if the image was pasted in an unnatural location.
On the other hand, an image in which the shape of the cornea in the combined panoramic image was kept circular, the center was not missing, and there were no unnaturally pasted parts was classified as a "success". The number of images in which either the right- or left-rotated part of the image was correctly merged was 75. 
\fgref{result_ics} compares the results of the manual and ICS syntheses. Among the videos that failed to be synthesized using ICS, those with distorted shapes or unnatural pasting were verified to contain frames that deviated from the cornea owing to contamination on the specular microscope or a shaking camera. It is considered that the stains themselves became feature points, and the pasting of the composite part failed. Additionally, the position of the image was significantly changed because of significant blurring, resulting in an unnatural position for pasting and distorting the overall shape of the image.

The results of the ICS and manually composited images were almost identical, suggesting that there was no problem with the image-compositing algorithm and that the mouse cornea was not captured in the first place when the specular microscope was used. The image was taken from the center of the mouse cornea in an upward direction, followed by leftward rotation along the edge of the cornea, and then a downward direction was taken at the point where the image had gone around the edge. If there is an area in the center of the cornea that has not been photographed at this time, the image will be missing.

\begin{table}[tb]
	\begin{center}
	\caption{Synthesis results with ICS}
\begin{tabular}{l|r|r}
\hline
Result          & Left & Right   \\ \hline
\hline
image dropout   & 15   & 20                                     \\
distorted shape & 6    & 12                                     \\
unnatural paste & 10   & 6                                     \\ \hline
\hline
success         & 63   & 56   
\\ \hline
\end{tabular}
	\label{tb:result_ICS}
	\end{center}
\end{table}

\subsection{Segmentation of guttae in panoramic images}
\begin{figure}[t]
\begin{center}
\includegraphics[width=\linewidth]{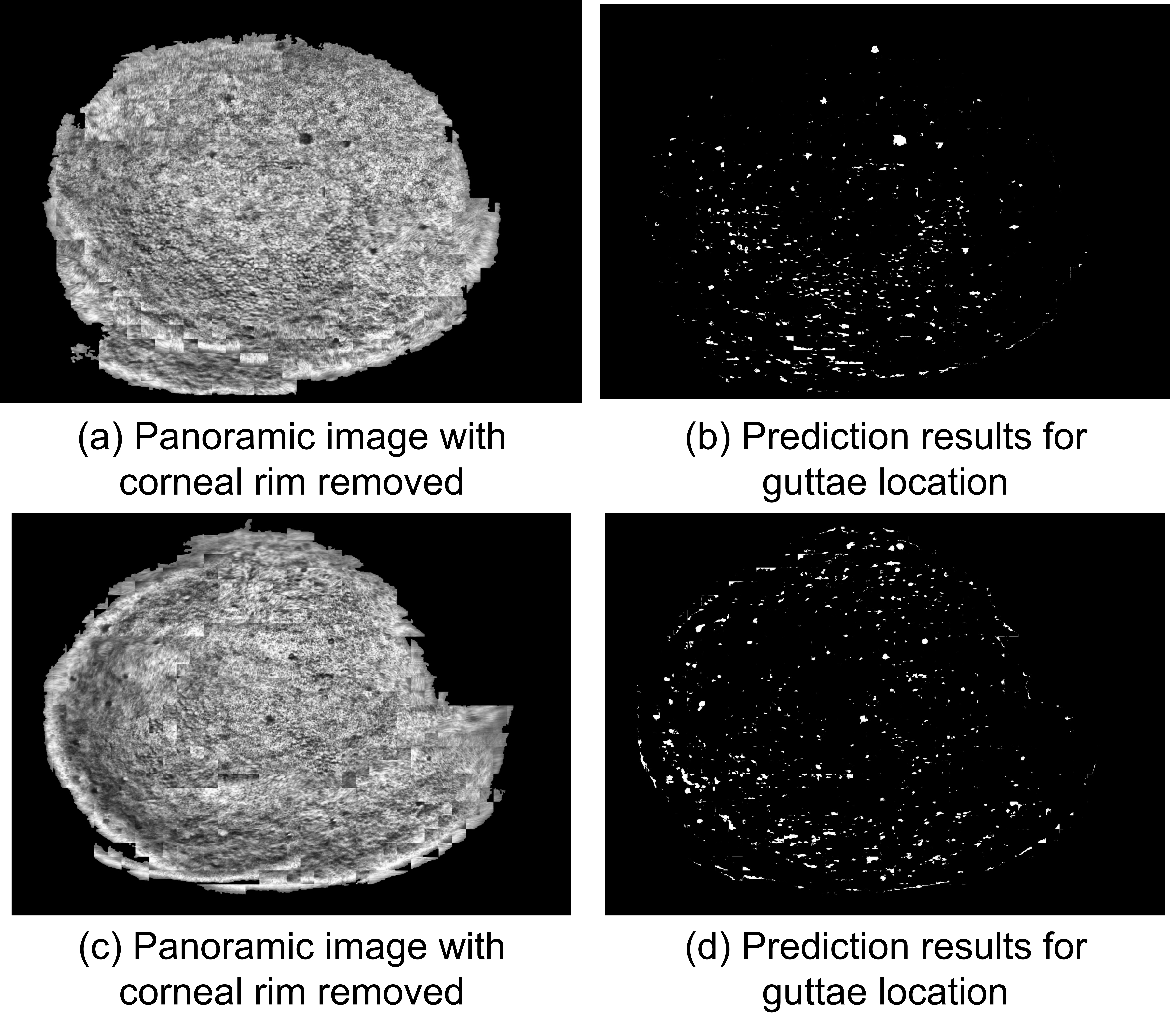}
\caption{\label{result_guttae_predict} Two examples of panoramic images and guttae locations.}
\end{center}
\end{figure}

The model that detects the guttae location was applied to the panoramic images of the videos (\fgref{framework_overview}, Dataset 1). 
\fgref{result_guttae_predict} shows two prediction examples of the guttae position in a panoramic image: (c) and (d) are the prediction results in (a) and (b), respectively. The original panoramic images were combined using ICS, and the edges of the corneas were cropped after sharpening.

Segmentation was performed on the images synthesized from the entire cornea using ICS. Currently, 21 still images and a mask image annotated with guttae are used for segmentation into training and validation datasets. The model was evaluated by comparing the validation data with predicted results.
\fgref{result_guttae_predict} shows that while the segmentation of the likely guttae is successful, it also partially predicts the guttae at the edges of the cornea. Further quantitative evaluation is essential; however, it may pose some limitations for future research. For this purpose, it is necessary to prepare an image to which the Grand Truth of the guttae is assigned. 

\subsection{Discussion}
It was observed that the algorithm implemented in OpenCV could not correctly synthesize results using mouse corneal endothelial cell images. The results showed that the images were pasted in unnatural positions compared to those obtained using artificial cornea data. The significant difference between the mouse corneal endothelial cell image data and the artificial cornea data is that, with the artificial cornea data, all images are in focus and the images themselves are not blurred. For the image data of mouse corneal endothelial cells, the process of extracting images in focus involved extracting images at equal intervals in chronological order, which resulted in the extraction of out-of-focus images. In contrast, the ICS-based method synthesized the images more accurately than the OpenCV-based synthesis software did because ICS uses a feature extraction method appropriate for corneal endothelial cells.

By contrast, OpenCV synthesis uses SIFT for feature extraction. This synthesis is considered to be progressing well. Until now, it has not been possible to obtain images of the entire cornea because of the narrow imaging range of specular microscopy. This study suggests that it is possible to obtain a composite image of the entire cornea by extracting a relatively focused image from a video of the entire cornea.

\section{Conclusions and Future Work}
The status of the entire cornea is currently inferred from the center of diagnosis and observation of corneal endothelial cells using specular microscopy. If images of the entire cornea could be obtained, more studies would be possible. In this study, we proposed a framework for generating images of the entire cornea from videos captured using contact specular microscopy. Focused images were extracted from the video and a panoramic composite image was generated. Furthermore, we constructed a learning model, U-Net, to extract the guttae from the entire image. To study the effectiveness of the proposed framework, we implemented it and applied it to corneal data from a mouse model of FECD. The panorama synthesis application used in the implementation was our custom-built ICS and the OpenCV algorithm, which is an open-source software. Artificial corneal images were synthesized with no unnatural aspects in the results. However, some of the extracted images were not correctly synthesized if they contained blurred images, and many images were correctly synthesized using ICS.

After the panorama was merged, the image was divided into a grid. Majority of the in-focus images were extracted and pasted, resulting in a sharper image than the previous output obtained using ICS. Using the extracted images within the region, we could also predict the guttae location. Although the implementation and application of the method to the data in this study confirmed its effectiveness, few quantitative evaluations have been performed. Quantitative evaluation, such as the accuracy of implementation, is an issue for the future.

%%%%%%%%%%%%%%%%% BIBLIOGRAPHY IN THE LaTeX file !!!!! %%%%%%%%%%%%%%%%%%%%%%

\bibliographystyle{unsrt}
\bibliography{AROB2023_ynagira}

\end{document}